\documentclass[twocolumn]{aastex701}
\usepackage{amsmath}
\begin{document}
\title{JWST Advanced Deep Extragalactic Survey (JADES) Data Release 5: Wisp Subtraction with the Non-negative Matrix Factorization Algorithm}

\author[0000-0002-8876-5248]{Zihao Wu}
\affiliation{Center for Astrophysics $|$ Harvard \& Smithsonian, 60 Garden St., Cambridge MA 02138 USA}
\email[show]{zihao.wu@cfa.harvard.edu}

\author[orcid=0000-0002-9280-7594]{Benjamin D.\ Johnson}
\affiliation{Center for Astrophysics $|$ Harvard \& Smithsonian, 60 Garden St., Cambridge MA 02138 USA}
\email{benjamin.johnson@cfa.harvard.edu}  

\author[orcid=0000-0002-2929-3121]{Daniel J.\ Eisenstein}
\affiliation{Center for Astrophysics $|$ Harvard \& Smithsonian, 60 Garden St., Cambridge MA 02138 USA}
\email{deisenstein@cfa.harvard.edu}  

\author[orcid=0000-0002-1617-8917]{Phillip Cargile}
\affiliation{Center for Astrophysics $|$ Harvard \& Smithsonian, 60 Garden St., Cambridge MA 02138 USA}
\email{pcargile@cfa.harvard.edu}

\author[orcid=0000-0003-4565-8239]{Kevin Hainline}
\affiliation{Steward Observatory, University of Arizona, 933 N. Cherry Avenue, Tucson, AZ 85721, USA}
\email{kevinhainline@arizona.edu} 

\author[orcid=0000-0002-8543-761X]{Ryan Hausen}
\affiliation{Department of Physics and Astronomy, The Johns Hopkins University, 3400 N. Charles St., Baltimore, MD 21218}
\email{rhausen@ucsc.edu}

\author[orcid=0000-0002-5104-8245]{Pierluigi Rinaldi}
\affiliation{Space Telescope Science Institute, 3700 San Martin Drive, Baltimore, Maryland 21218, USA}
\email{prinaldi@stsci.edu} 

\author[orcid=0000-0002-4271-0364]{Brant E. Robertson}
\affiliation{Department of Astronomy and Astrophysics, University of California, Santa Cruz, 1156 High Street, Santa Cruz, CA 95064, USA}
\email{brant@ucsc.edu}  

\author[orcid=0000-0002-8224-4505]{Sandro Tacchella}
\affiliation{Kavli Institute for Cosmology, University of Cambridge, Madingley Road, Cambridge, CB3 0HA, UK}
\affiliation{Cavendish Laboratory, University of Cambridge, 19 JJ Thomson Avenue, Cambridge, CB3 0HE, UK}
\email{st578@cam.ac.uk}  

\author[orcid=0000-0003-2919-7495]{Christina C. Williams}
\affiliation{NSF National Optical-Infrared Astronomy Research Laboratory, 950 North Cherry Avenue, Tucson, AZ 85719, USA}
\email{christina.williams@noirlab.edu} 

\author[orcid=0000-0001-9262-9997]{Christopher N. A. Willmer}
\affiliation{Steward Observatory, University of Arizona, 933 N. Cherry Avenue, Tucson, AZ 85721, USA}
\email{cnaw@as.arizona.edu} 

\begin{abstract}
Wisps are among the most prominent scattered light artifacts in JWST/NIRCam imaging. They often appear in certain regions of the detectors and contaminate observations at surface-brightness levels relevant for faint-source photometry. We introduce a new subtraction method that uses the non-negative matrix factorization (NMF) algorithm to model and remove wisps. Using deep NIRCam observations from the JWST Advanced Deep Extragalactic Survey (JADES) and other programs, we construct multi-component, filter- and detector-specific wisp templates that capture the wisp structures and their exposure-to-exposure morphological variations. Wisps in individual exposures are represented as non-negative linear combinations of these templates, consistent with their additive nature and reducing degeneracies relative to single-template scaling. Compared to existing approaches, our method delivers lower residual root mean square in wisp-affected regions and reduces photometric bias and scatter to levels nearly consistent with clean detector areas. The NMF wisp templates are readily applicable to other datasets and are publicly released to support future NIRCam extragalactic surveys.
\end{abstract}

\section{Introduction}
\setcounter{footnote}{0}
\begin{figure*}
    \centering
    \includegraphics[width=0.95\linewidth]{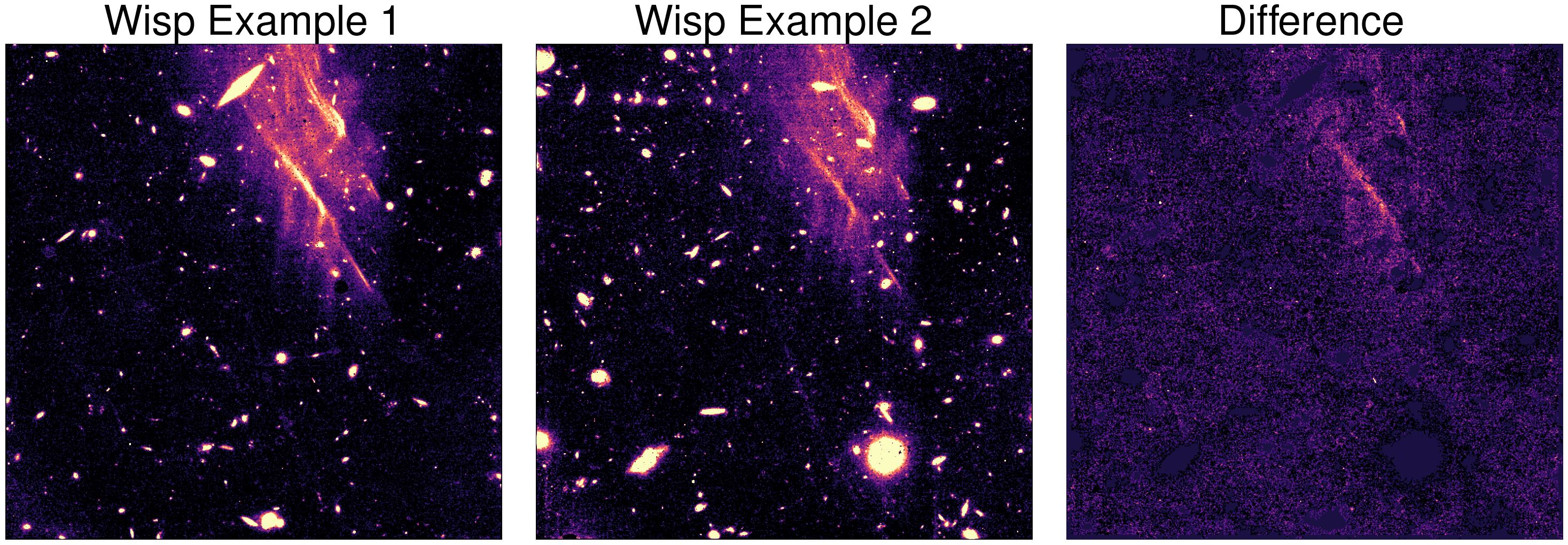}
    \caption{The left and middle panels are two examples of wisps in the NIRCam B4 detector in the F150W band from different exposures. The right panel shows their difference, demonstrating exposure-to-exposure morphological variations, especially in the relative strengths of wisp substructures. Other sources are masked in the image differencing.}
    \label{fig:wisp_example}
\end{figure*}

JWST has been revolutionizing astronomy with its exceptional imaging capabilities \citep{Rieke2023JWST}. Yet, as with any state-of-the-art instrument, JWST has several artifacts due to unexpected scattered light \citep{Rigby2023jwst}.  Wisps are among the most prominent artifacts in the JWST NIRCam short-wavelength (SW; 0.6--2.3\,$\mu$m) detectors, which affect a large area of the field of view \citep{Bagley2023ceers, Eisenstein2023JADES, Franco2025cosmoswebb}. Removing the wisp artifacts is essential for precise photometric and morphological analysis.

Wisps are caused by faint, diffuse scattered light that enters the telescope through a rogue path. They are likely caused by off-axis light that scatters off the secondary mirror support strut and illuminate the detector  \citep{Rigby2023jwst}. The illuminating source seems to appear at a fixed position, independent of telescope pointing or orientation over a wide range of configurations \citep{Sunnquist2022report, Rigby2023jwst}. As a result, wisps constantly appear in certain regions of the NIRCam detectors with similar morphology \citep{Sunnquist2022report, Eisenstein2023JADES}. 

The wisp morphology has slight but visible variations in different observations (see examples in Figure \ref{fig:wisp_example}). The morphology variation makes it difficult to subtract using a single  templates. The wisp morphologies change with filters, and the amplitude increases toward longer wavelengths in the SW channel. In the F200W band, the brightest regions reach 0.05\,MJy\,sr$^{-1}$, a significant contamination for sources fainter than 25\,AB\,mag  \citep{Sunnquist2022report, Eisenstein2023JADES, robotham_2023_wisp}.

Two major subtraction methods have been developed. The Space Telescope Science Institute (STScI) pipeline uses the median wisp templates to subtract wisps\footnote{https://jwst-docs.stsci.edu}: they construct wisp templates by median-stacking a large number of public NIRCam images; these templates are then scaled to match the observed strength of each exposure and subtracted from the images. Similar techniques have been developed and applied independently, e.g., by the COSMOS-Web team \citep{Franco2025cosmoswebb}. An alternative method removes wisps dynamically by comparing SW and long-wavelength (LW) images \citep{robotham_2023_wisp}. They construct wisp templates individually for each exposure by subtracting and interpolating all sources in the image, where source morphologies and fluxes are inferred from LW images that are free from wisp contamination.

The two methods differ in their strengths with respect to sensitivity and flexibility. The template-based approach is more sensitive to diffuse structures, as stacking numerous exposures enhances the signal-to-noise ratio (SNR). However, the single template lacks the flexibility to account for morphological variations, often leaving noticeable residuals when the wisp morphology deviates from the template. In contrast, the dynamic method allows for arbitrary morphologies. However, the SNR in single exposures is usually not sufficient to capture diffuse wisp features. Moreover, this method relies on accurate source subtraction, which may introduce artifacts if source morphologies vary across wavelengths or if sources appear in SW images but are not detected in the LW data \citep{robotham_2023_wisp}.

The existing wisp-subtraction approaches face a fundamental trade-off between sensitivity and flexibility: median-based templates achieve high signal-to-noise for diffuse structures but cannot accommodate exposure-to-exposure morphological variations, while dynamic, exposure-level methods allow arbitrary morphologies but are limited by noise and uncertainties in source subtraction. This work resolves this trade-off by combining the sensitivity of deep, stacked data with the flexibility of a multi-component, non-negative template basis. We construct the templates using a non-negative matrix factorization (NMF) algorithm \citep{lee1999Natur}, which are optimized using extensive deep NIRCam images in the JWST Advanced Deep Extragalactic Survey (JADES; \citealt{rieke2020a, bunker2020a, Eisenstein2023JADES}).  This method has been incorporated into the JADES data reduction pipeline, where it provides more accurate subtraction of wisps compared to existing approaches \citep{Johnson2026}.

The paper is structured as follows. We describe the algorithm for template construction in Section~\ref{sec:formalism} and implementation in Section~\ref{sec:implementations}. We describe the use of the templates to subtract wisps in Section~\ref{sec:subtraction}. We present the results and comparison with other methods in Section~\ref{sec:results}. We discuss caveats in Section~\ref{sec:discussion} and summarize in Section~\ref{sec:summary}. In Appendix \ref{sec:skyflat} we explore the impact of flat field errors and assess the quality of the current flat fields.

\begin{figure*}
    \centering
    \includegraphics[width=0.95\linewidth]{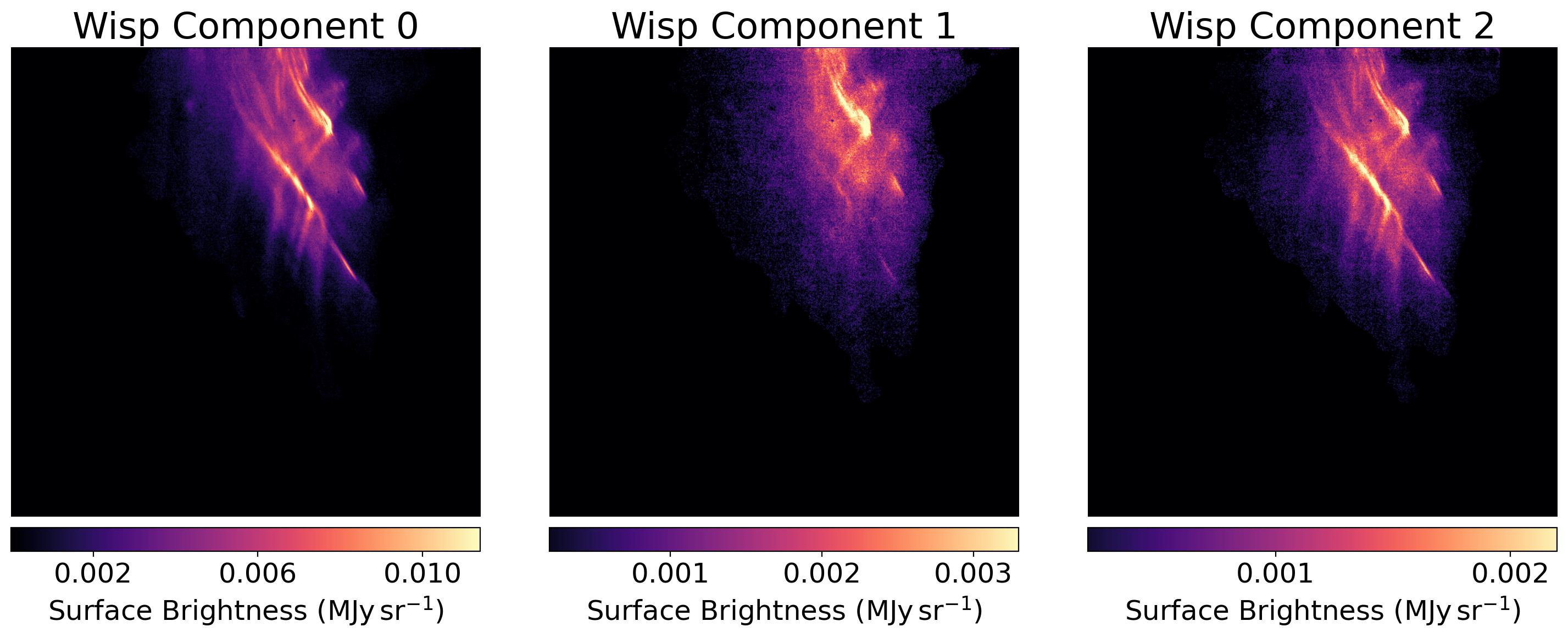}
    \caption{Multi-component wisp templates derived from the weighted non-negative matrix factorization (NMF) algorithm for the F150W band in the NIRCam B4 detector. The three components are combined to capture the dominant wisp morphology and its exposure-to-exposure variation. From left to right, the contribution to the wisp flux decreases. The template brightness reflects the typical flux level. For reference, $0.01\,\mathrm{MJy\,sr^{-1}}$ is $25.5\,\mathrm{mag\,arcsec^{-2}}$, and the width of the NIRCam B4 detector corresponds to 64\,arcmin.}
    \label{fig:wisp_comp}
\end{figure*}

\section{Formalism}
\label{sec:formalism}
This section introduces the NMF algorithm used to construct the multi-component wisp templates. Our goal is to derive a set of wisp templates that optimally minimize the residuals across the full dataset. Formally, this can be posed as a matrix approximation problem: given the data matrix $\mathbf{D} \in \mathbb{R}^{m \times n}$, representing $m$ exposures each with $n$ pixels, we seek wisp templates $\mathbf{W}$ and wisp amplitudes $\mathbf{A}$ that minimize the loss function:

\begin{equation}
\min_{\mathbf{A}, \mathbf{W} \geq 0}  \mathcal{L}(\mathbf{A}, \mathbf{W}) = \frac{1}{2} \|\mathbf{D} - \mathbf{A} \mathbf{W} \|^2.
\end{equation}
For a set of $k$ templates, we define $\mathbf{W}\in\mathbb{R}_{\geq0}^{k \times n}$ and $\mathbf{A}\in\mathbb{R}_{\geq0}^{m \times k}$.  The $\|\cdot\|$ operator denotes the Frobenius norm weighted by the inverse of noise variance of each pixel, denoted as $\mathbf{\Omega}\in \mathbb{R}_{\geq 0}^{m \times n}$. Pixel masks are implemented by setting the corresponding elements of $\mathbf{\Omega}$ to zero. Details of the masking procedure are described in Section~\ref{sec:implementations}.

The loss function in the element-wise format is:
\begin{equation}
\mathcal{L}(\mathbf{A}, \mathbf{W}) = \frac{1}{2}\sum_{m,n}\Omega_{mn}  \left( D_{mn} - \sum_k A_{mk} W_{kn} \right)^2.
\end{equation}
At the optimal solution, the gradients of $\mathcal{L}$ with respect to each element of $\mathbf{A}$ and $\mathbf{W}$ vanishes:
\begin{align}
\frac{\partial \mathcal{L}}{\partial A_{mk}} &= - \sum_n \Omega_{mn} W_{kn}  \left( D_{mn} - \sum_l A_{ml} W_{ln} \right)  = 0, \label{eq:minimal1}\\
   \frac{\partial \mathcal{L}}{\partial W_{kn}} &= - \sum_i \Omega_{mn} A_{mk} \left( D_{mn} - \sum_l A_{ml} W_{ln} \right) = 0. \label{eq:minimal2}
\end{align}
The NMF algorithm achieves the minimal by iteratively updating $\mathbf{W}$ and $\mathbf{A}$ in the following format  \citep{lee2000algorithms, zhu_2016_nmf, Ren2018NMF}

\begin{align}
W_{kj} \leftarrow W_{kj} \cdot \frac{\sum_i A_{ik} \Omega_{ij} D_{ij}}{\sum_i A_{ik} \Omega_{ij} (AW)_{ij}},  \\
A_{ik} \leftarrow A_{ik} \cdot \frac{\sum_j W_{kj} \Omega_{ij} D_{ij}}{\sum_j W_{kj} \Omega_{ij} (AW)_{ij}}.
\end{align}
In matrix notation, the update rule is:
\begin{align}
\mathbf{W} \leftarrow \mathbf{W} \circ \frac{\mathbf{A}^{\top} (\mathbf{\Omega} \circ \mathbf{D})}{\mathbf{A}^{\top} (\mathbf{\Omega} \circ (\mathbf{A} \mathbf{W}))}, \label{eq:updates1}\\
\mathbf{A} \leftarrow \mathbf{A} \circ \frac{(\mathbf{\Omega} \circ \mathbf{D}) \mathbf{W}^{\top}}{(\mathbf{\Omega} \circ (\mathbf{A} \mathbf{W})) \mathbf{W}^{\top}}, \label{eq:updates2}
\end{align}
where $\circ$ denotes element-wise multiplication and the division is performed element-wise.

This update rule satisfies Equations~\ref{eq:minimal1} and \ref{eq:minimal2} after convergence, while maintaining the non-negativity of each $\mathbf{W}$ and $\mathbf{A}$ element. The update rule is intuitive: in Equation~\ref{eq:updates1}, the templates are updated by the ratio of the data $\mathbf{D}$ to the wisp model $\mathbf{A}\mathbf{W}$, weighted by the noise matrix $\mathbf{\Omega}$ and the amplitudes $\mathbf{A}$. Each pixel is updated independently, rescaling the wisp model toward the data. Similarly, in Equation~\ref{eq:updates2}, the amplitudes are updated by the data-to-model ratio weighted by the noise and the template flux. The template matrix $\mathbf{W}$ and the amplitude matrix $\mathbf{A}$ are updated iteratively until convergence.

The NMF method decomposes structures into a small number of prominent basis components. Its objective is similar to that of principal component analysis (PCA), but with the key distinction that all elements of $\mathbf{W}$ and $\mathbf{A}$ are constrained to be non-negative. This constraint is essential not only for reducing model degeneracy and overfitting, but also for maintaining physical consistency with the additive nature of wisp contamination. We also explored the PCA-based approach but found that it performs significantly worse than the NMF method, as discussed in Section~\ref{sec:discussion}.

The NMF algorithm is one of the most effective strategies for solving the non-negative matrix factorization problem and has been widely applied across astrophysical contexts, including galaxy spectral decomposition \citep{Hurley2014MNRAS}, photometric redshift calibration \citep{Zhang2017ApJ}, and circumstellar disk modeling \citep{Ren2018NMF}. However, NMF guarantees convergence only to a local minimum. A global-minimum optimal solution is not guaranteed, as the problem is NP-hard and no polynomial-time algorithm is known for any such problems \citep{Ding_2005_optimal, vavasis2010complexity}. As a result, the choice of initialization and optimization strategy plays a critical role in achieving robust performance. We describe our implementation in the next section.

\section{Wisp Template Construction}
\label{sec:implementations}
This section describes the procedures for building wisp templates using the NMF algorithm.
\subsection{Data and Pre-process}
\label{sec:preprocess}
We fit wisp templates from NIRCam images in the wide-band filters F090W, F115W, F150W, F200W and the medium-band filters F162M, F182M, F210M. Exposure time of each band is listed in Table~\ref{tab:exposures}. The dataset includes observations from JADES \citep{Eisenstein2023JADES} and programs 1283 \citep{Perez_Gonzalez2024ApJ, Ostlin2025A&A}, 1895 (\citealt{Oesch2023FRESCO}), 1963 (\citealt{Williams2023JEMS}), 2514 (\citealt{Williams2025Panoramic}), 2516 (\citealt{Hodge2025survey}), 3215 (\citealt{Eisenstein2025ApJS}) 3577 (PI: Egami), 3990 (\citealt{Morishita2025survey}), 5997 (Looser et al. in prep.), 6541 (PI: Egami). The data are reduced with the JADES pipeline \citep{Johnson2026}. Wisp templates are constructed independently for each band. We do not build templates for the F070W band, as no significant wisp features are present. 

\begin{deluxetable*}{lccccccc}
\tablecaption{Data summary \label{tab:exposures}}
\tablewidth{0pt}
\tablehead{
  \colhead{} &
  \colhead{F090W} &
  \colhead{F115W} &
  \colhead{F150W} &
  \colhead{F162M} &
  \colhead{F182M} &
  \colhead{F200W} &
  \colhead{F210M}
}
\startdata
$N_{\rm exp}$      & 404 & 667 & 379 & 45  & 202 & 361 & 133 \\
$t_{\rm exp}$ (ksec) & 452.9 & 925.2 & 484.3 & 123.7 & 280.5 & 388.8 & 207.9 \\
\enddata
\tablecomments{NIRCam imaging used for wisp template fitting. For each filter, we list the number of individual exposures and the total exposure time in ksec. }
\end{deluxetable*}

We start with the {\tt stage 1} results from the JADES pipeline that uses context map {\tt jwst\_1228.pmap} \citep{Johnson2026}, which are detector-level count-rate images. Since wisp subtraction is performed in the subsequent {\tt stage 2} calibration, these represent the latest data that retain the raw wisp signal. Prior to wisp fitting, we pre-process the images by performing flat-field calibration, background subtraction, hot-pixel removal, and $1/f$ noise correction. These steps largely follow the JADES  pipeline, but we adopt more aggressive outlier-rejection criteria to improve the robustness of the wisp fitting. To accelerate the fitting, we downsample the data by block-averaging over $4\times4$ pixel bins and revert to the full-resolution data at the final stage. Below, we detail these procedures.

We mask sources and other artifacts to prevent contamination of the wisp fitting. The identification and masking procedures are described in \cite{Johnson2026}. Briefly, sources are masked according to the segmentation maps derived from the detection images that stack LW images \citep{Robertson2026}, and artifacts are masked through visual inspection by the JADES team \citep{Johnson2026}. For regions not covered by the segmentation maps, we construct additional masks by stacking all LW images within the same association of image program ID, position angle, epoch, and general sky region.  All masks are dilated by eight iterations to capture extended features beyond the original segmentation boundaries.

We perform flat field calibration for the {\tt stage 1} data products using the STScI flat fields. We subtract the median background for each exposure image. During background estimation, sources, artifacts, and wisps are masked, where wisp masks are derived from the STScI wisp templates. We correct for $1/f$ noise using a simple column-based median subtraction\footnote{https://github.com/chriswillott/jwst}, in which the channel-wise median background is removed. We use the {\tt Bottleneck} package to accelerate the median estimation. This method is fast and sufficient for wisp template construction. Because wisps can span nearly an entire channel, masking them may leave too few pixels to reliably estimate the $1/f$ noise. To mitigate this issue, we estimate the $1/f$ noise after subtracting the STScI wisp templates, instead of masking the wisps. This subtraction is only used for $1/f$ noise estimate, and we do not subtract wisps from the original image since we will model the wisps. Finally, we remove bad pixels with two procedures: we first apply a median filter with a box size of seven pixels to individual exposures and reject pixels that are $10\,\sigma$ outliers compared with their {\tt error} map; we then apply a median filter to the mean stack images and mask pixels in all exposures for $10\,\sigma$ outliers. This outlier rejection is more aggressive than that used in the early stages of the JADES pipeline, but it significantly improves the quality of the wisp templates by removing nearly all hot pixels that would otherwise leave imprints in the templates.

\subsection{Building the Wisp Templates}
\label{sec:make_template}
We construct multi-component wisp templates using the NMF algorithm applied to the pre-processed data. We first identify wisp regions from the mean-stacked images for each filter and detector. These regions define the detector areas with significant wisp contamination where the fitting is performed. We generate segmentation maps and manually adjust the threshold to ensure that all extended structures associated with the wisps are included. Only pixels within the identified wisp regions are used in the NMF optimization, which improves computational efficiency and reduces contamination from unrelated artifacts.

We exclude exposures with significant contaminants inside the wisp region. The contaminants include diffraction spikes of stars and luminous galaxies with diffuse emission that are not fully masked. We identify these contaminants by performing singular value decomposition (SVD) of the data and visually inspecting the first nine components. These contaminants show up in the SVD components, and we exclude the associated exposures.

We implement the NMF algorithm to construct the optimal wisp templates. As the NMF algorithm does not guarantee a global minimum, the initialization is important. Motivated by the sequential NMF process in \cite{Ren2018NMF}, we start by fitting only one template, and gradually add more components to the NMF optimization, where previous results are used as the initial values for the next optimization. Specifically, the templates from the $k$-component optimizations are used as the initial step for the first $k$-components for optimization with $k+1$ components. We find it a good practice to downscale the amplitudes to 80\% in the initialization, as the full amplitudes can enforce a large fraction of image to negative residuals, which hinders the optimization of the new component due to the non-negative constraint. We initialize the first component as the mean-stacked image and initialize each new component by fitting it to the residual obtained after subtracting the downscaled $k$-component model. In the NMF optimization, we find it improve stability by constraining the multiplicative update factor between 0.1 and 10 in each iteration. While the NMF algorithm does not have a strict stopping criterion, we find 300 iterations are usually sufficient for convergence.

We find that three-component templates provide the most effective wisp subtraction for the A3 and B4 detectors in each band, while a single template is sufficient for the A4 and B3 detectors. Adding additional components does not extract meaningful wisp structure. Instead, higher-order templates appear much fainter and are dominated by spurious artifacts, such as hot pixels, $1/f$ noise, and other scattered-light features. This behavior suggests that we have reached the maximum useful set of templates, as no valuable information remains in the data.

Finally, we use the optimized templates and amplitudes derived from the downsampled data to initialize the optimization on the full-resolution images. We first refine the templates for 100 iterations with the amplitudes held fixed, and then jointly optimize both templates and amplitudes for an additional 100 iterations. To further mitigate residual $1/f$ noise, we alternate between wisp optimization and $1/f$ noise correction on the wisp-subtracted images, repeating this procedure three times. As an example, the final wisp templates for the F150W band are shown in Figure~\ref{fig:wisp_comp}. Templates for other bands and detectors are also publicly available\footnote{https://github.com/zihaowu-astro/NMFwisp}.

Wisps in the NIRCam A3 detector are located in a region with persistence artifacts, which is well-known from laboratory experiments \citep{Leisenring2016persistence}. The persistence is caused by sky backgrounds with consistent illumination patterns. Different from persistence caused by stars, the background persistence is diffuse and appears with consistent morphology, making it possible to subtract using templates. We construct the persistence template by median-stacking F070W images, where wisps are absent. While the persistence morphology may evolve over time due to spatial variations in trap density and decay timescales, constraining such evolution is beyond the scope of this work. We therefore adopt the fixed persistence template as an approximation. Persistence and wisps are then fit simultaneously by including persistence as an additional component in the NMF model, with only its amplitude allowed to vary. We do not optimize the persistence morphology because it is highly degenerate with the wisp structures.

The non-negativity constraint in the NMF algorithm introduces a mild bias in noise-dominated regions. While wisp emission is intrinsically non-negative, noise fluctuations can take both positive and negative values. By construction, NMF suppresses pixels whose summed signal across exposures is negative, forcing the corresponding model values to zero. This asymmetric treatment of positive and negative fluctuations leads to a statistical bias. The resulting bias is of order the noise divided by the square root of the number of images used for template construction. With $\sim$ 300 exposures, this bias is about $6\%$ of the noise. We estimate the bias distribution using the mean-stacked data after wisp subtraction and then subtract it from individual exposures. After this adjustment, the residual systematic uncertainty is below 1\% of the noise for typical wisp amplitudes. The bias only affects only a thin layer at the wisp outskirts since our wisp templates are defined within regions showing significant wisps.

\begin{figure*}
    \centering
\includegraphics[width=0.98\linewidth]{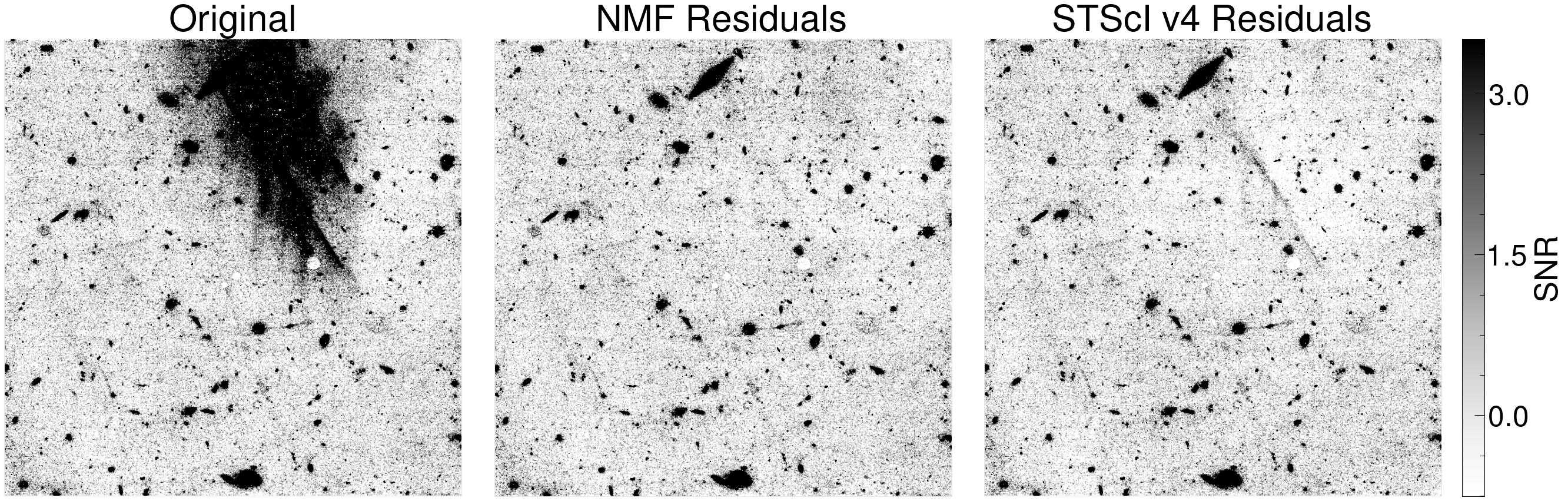}
    \caption{Example NIRCam F150W image in the B4 detector. The left panel shows the original image, the middle panel shows the result after wisp subtraction using the NMF templates, and the right panel shows the result using the STScI v4 templates. The color bar indicates the signal-to-noise ratio (SNR).}
    \label{fig:residual_single}
\end{figure*}

\begin{figure*}
    \centering
    \includegraphics[width=0.98\linewidth]{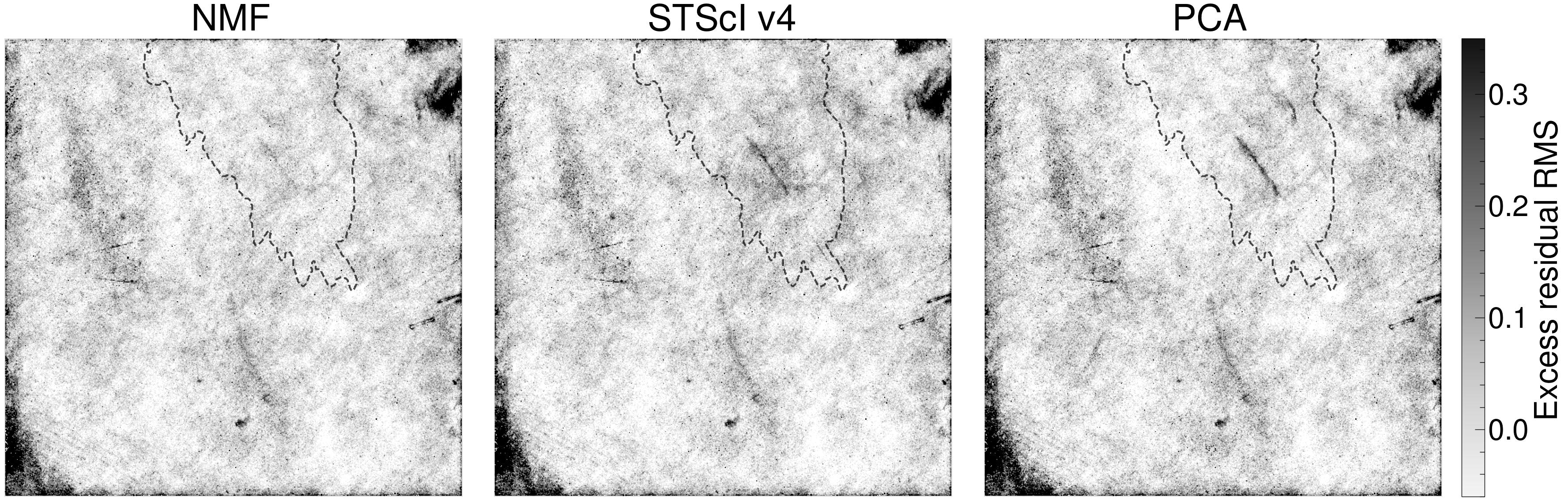}
    \caption{Comparison of the excess residual root mean square (RMS) in the F150W band for the NIRCam B4 detector after wisp subtraction using the NMF-based, STScI v4, and PCA templates. The color bar indicates the excess residual RMS, which quantifies errors arising from effects beyond background noise. It is defined as the difference between the measured RMS across the dataset and the RMS expected from background noise based on the formal errors, normalized by the noise level. The dashed contour delineates regions with prominent wisps. Within this contour, the mean excess RMS obtained with the NMF method is $\sim$50\% of that from the STScI method, while the sharpest residual features are only $\sim$30\% of the STScI excess RMS. Outside the contour, most regions are relatively clean; the structures are primarily associated with persistence, unmasked stellar diffraction spikes, and other scattered-light artifacts. } 
    \label{fig:residual_all}
\end{figure*}

\section{Wisp Subtraction}
\label{sec:subtraction}
We describe our strategy for subtracting wisps using the templates constructed in Section~\ref{sec:make_template}. We adopt the non-negative least squares (NNLS) algorithm \citep{lawson1995solving}, as implemented in {\tt SciPy}, to determine the amplitudes of each wisp component. The $\chi^2$ is weighted by the image uncertainties by scaling both the data and the templates by the errors in the NNLS optimization. Sources and other artifacts are masked during the fit following the procedure described in Section~\ref{sec:make_template}.

We find that the stability improves by fitting only in high-SNR wisp regions. The high-SNR regions are defined as the regions where the wisps are brighter than their standard deviation within the wisp masks. We provide masks for high-SNR wisp regions as an extension of the wisp template products. Fitting wisp amplitudes in the high-SNR regions is more sensitive to sharp wisp features and less susceptible to contamination from persistence and extended flux from galaxies. That said, the subtraction is performed over the entire wisp-contaminated area to subtract their diffuse flux.

We subtract wisps while iteratively correcting for $1/f$ noise because the two artifacts are degenerate. We adopt the column-based median approach for the $1/f$ noise correction, as in Section~\ref{sec:preprocess}, which is fast and sufficient for wisp estimation. The iteration needs to be repeated at least three times according to our experience. 

We propagate wisp uncertainties to the images, considering both wisp amplitude errors and imperfections in the wisp templates. The amplitude uncertainties are estimated using a generalized least-squares framework for individual wisp components, which accounts for data noise and model degeneracy. Template errors arise from exposure-to-exposure variations in wisp morphology that are not fully captured by the finite template basis. We estimate the template errors empirically as the pixel-wise standard deviation of the residual images after wisp subtraction across the full JADES dataset. The two sources of uncertainty are propagated in quadrature into the data error image.

\section{Results}
We find substantial improvements in wisp subtraction when using the multi-component templates derived from the NMF algorithm. Figure~\ref{fig:residual_single} shows an example from the NIRCam B4 detector in the F150W band, compared with results obtained using the STScI v4.0 template under the same reduction procedures, including background subtraction and $1/f$ noise correction.

We further quantify the performance through a statistical comparison by computing the root mean square (RMS) of the residuals across all exposures in the JADES dataset. As shown in Figure~\ref{fig:residual_all}, the multi-component templates yield residuals at a level comparable to the local background. Both sharp and extended wisp features are more effectively removed than with the STScI templates.

\begin{figure}
    \centering
\includegraphics[width=0.9\linewidth]{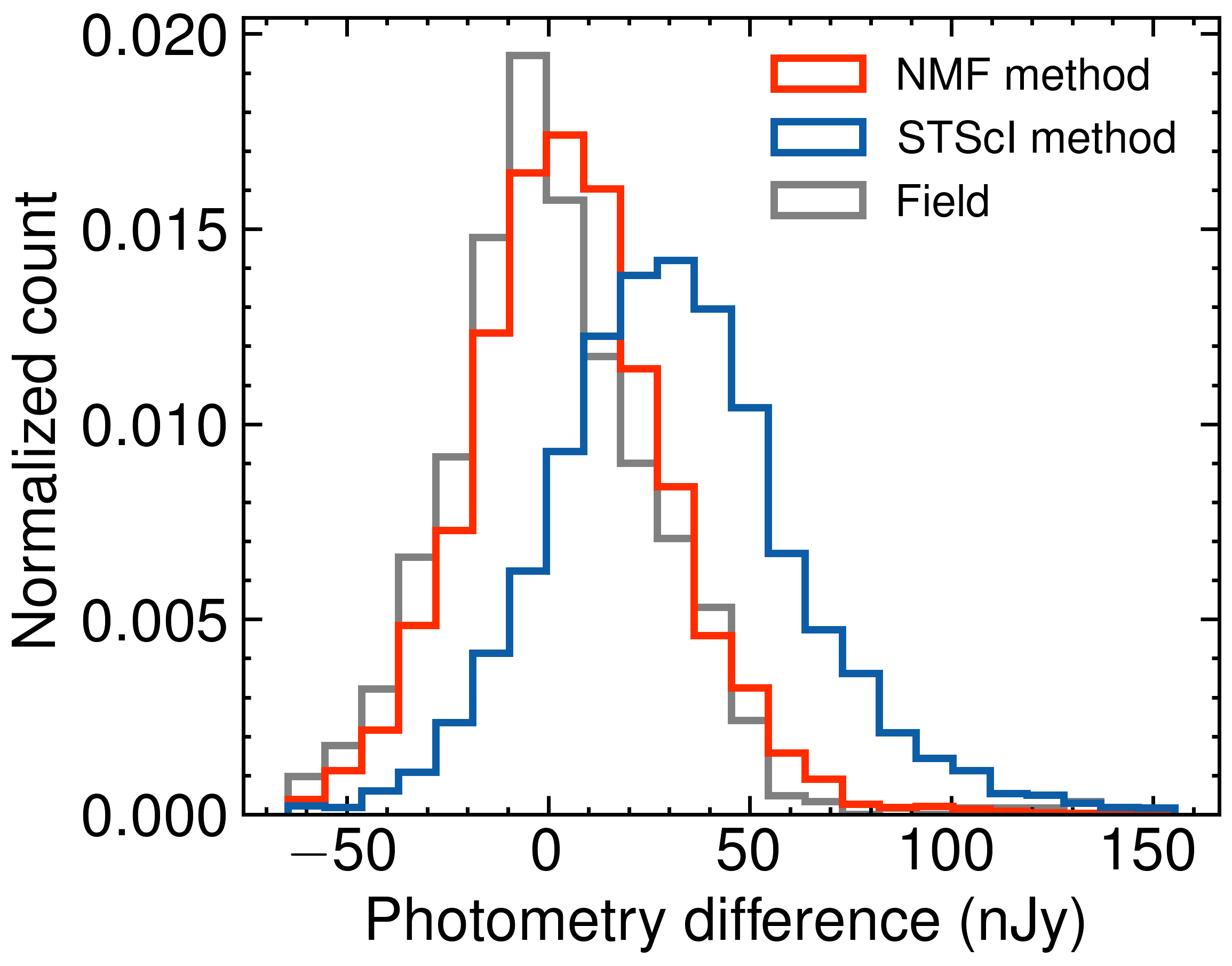}
    \caption{Photometric differences between wisp-affected and clean regions for the same sources. The photometry is performed with a 0.5$''$-radius aperture on exposure images after wisp subtraction, using either the NMF multi-component templates or the STScI wisp templates. Results are shown for the NIRCam B4 detector in the F150W band. The field sample shows photometric differences measured outside wisp affected regions. The comparison does not include local background subtraction.}
    \label{fig:photometry}
\end{figure}

We also experiment with constructing wisp templates using a noise-weighted PCA method \citep{Delchambre2015PCA}. However, NMF significantly outperforms PCA, even when using nine PCA components, as shown in Figure~\ref{fig:residual_all}. The difference in performance is likely because the non-negativity constraint in NMF ensures that it only captures additive light, consistent with wisps, whereas PCA is less physical and more susceptible to contamination from other effects.

We assess photometric errors introduced by residual wisps using a control experiment. We select sources that lie in wisp-affected regions in some exposures and in wisp-free regions in others. We perform aperture photometry with a 0.5$''$ radius on individual, wisp-subtracted exposures. For each source, we compare the flux measured in wisp-contaminated exposures to the reference flux, defined as the mean measured in clean regions. Figure~\ref{fig:photometry} shows the resulting flux differences for images processed with the NMF-based templates and the STScI templates. The NMF method yields both smaller bias and reduced scatter, with a distribution consistent with that of unaffected regions. However, we note that as we do not apply local background subtraction, the differences between the two methods are likely smaller when a two-dimensional background model is fit after wisp removal.

\begin{figure*}
    \centering
    \includegraphics[width=0.44\linewidth]{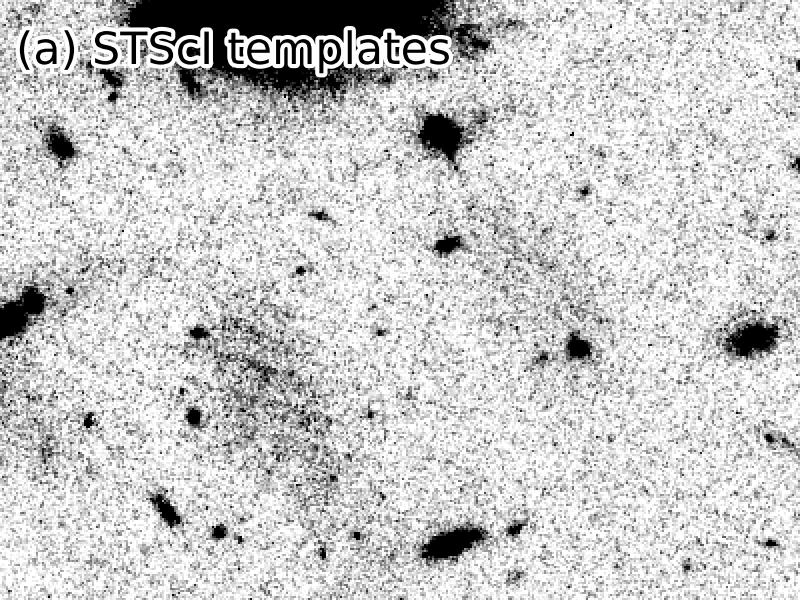}
    \includegraphics[width=0.44\linewidth]{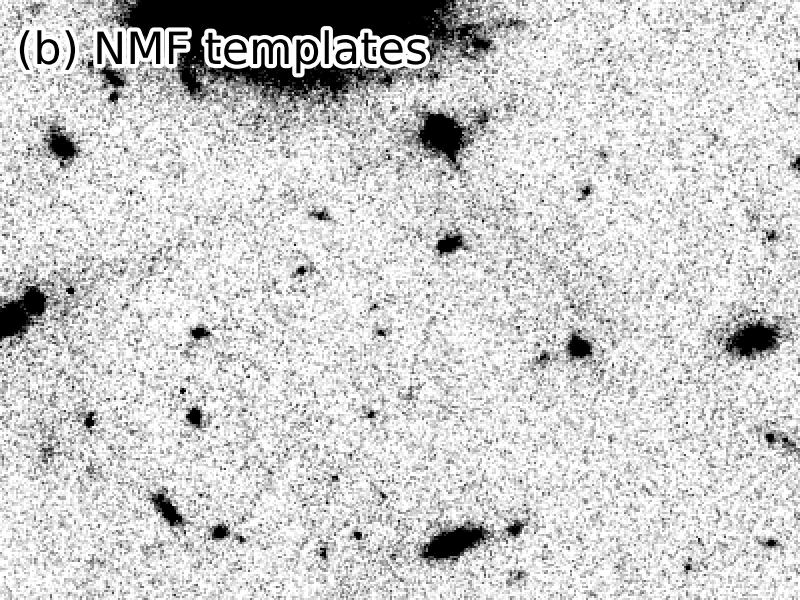}
    \caption{Example mosaic images combining multiple dithered exposures from the JWST program 1287. The left panel shows the mosaic after wisp subtraction using the STScI templates, while the right panel shows the result using the NMF templates. }
    \label{fig:mosaic}
\end{figure*}

Figure~\ref{fig:mosaic} shows an example mosaic image in the F200W band, demonstrating the improved mosaic quality achieved with our new wisp subtraction method. The STScI  templates leave residual wisps that persist even after stacking multiple exposures. In contrast, mosaics processed with our method show cleaner backgrounds and enhanced image quality, reducing source misidentification and improving the accuracy of photometric and morphological measurements.

\label{sec:results}

\section{Procedure for other programs}
The wisp templates are readily applicable to other observing programs. The wisp morphology is known to be stable across different fields of view and telescope position angles, and we do not observe significant differences between the GOODS-S and GOODS-N fields despite their wide separation. It is therefore unnecessary to construct new wisp templates for each program, and we recommend directly using our optimized templates. The only exception is the so-called ``bright wisps'', which exhibit distinct morphologies and are discussed in Section~\ref{sec:discussion}.

The wisp subtraction method can be integrated into the standard reduction pipeline by replacing the wisp-subtraction step in {\tt Stage~2}. Similar to the STScI method, sources must be masked prior to the wisp subtraction. Because NIRCam usually acquires SW and LW images simultaneously with identical fields of view, source detection can be performed using the LW images, where wisp contamination is absent.

The time cost of wisp subtraction is 0.4 second per exposure for a single NIRCam detector chip when using a single CPU core on the MacBook M4 Pro chip. The cost increases to 2 seconds when requiring joint fitting with the $1/f$ noise. We recommend the joint fitting, as $1/f$ noise can affect wisp inference. This $1/f$ noise correction is applied only internally during the wisp fitting. Because our correction scheme is simplified for computational efficiency, it is sufficient for robust wisp inference but may not be optimal for a clean noise subtraction. A more complete $1/f$ noise correction can be applied using the STScI pipeline after wisp subtraction.

Our code outputs the estimated wisp and its associated uncertainty, which includes contributions from the uncertainties in the fitted wisp amplitudes and in the wisp templates, as discussed in Section~\ref{sec:subtraction}.

\section{Discussion and Caveats}
\label{sec:discussion}
\subsection{Distribution of wisp amplitude}

The wisp flux is nearly constant across the JADES observations in both GOODS-S and GOODS-N fields. The wisp fluxes are difficult to directly estimate from data due to source contamination, so we measure the fluxes as those of our best-fit wisp model. We find the standard deviation of the flux is merely $\sim$ 12\%, indicating slight variation between exposures. The scatter is not likely caused by measurement errors, as our tests find that perturbing the wisp template amplitudes by as little as 5\% introduces noticeable residuals in the images.

The wisp flux shows only mild correlations between detectors. The Pearson correlation coefficients between adjacent detectors (A3--A4 or B3--B4) are $\sim$0.4, which is not statistically significant. We find no systematic trend with right ascension or declination, even across the large separation between the GOODS-S and GOODS-N fields, where both the morphology and amplitude of wisps remain similar. No correlation is found with observation date or telescope pointing. The relative strengths of the individual wisp templates show no significant dependence on either parameter.

\subsection{Bright wisps}
A distinct wisp-like stray-light feature, termed the              ``bright wisp'', has also been reported. These features do not appear in the JADES observations but have been seen in rare cases in other programs, such as the JWST Emission-Line Survey (PID 2321; \citealt{Duncan2025JELS}) and the SAPPHIRES program (PID 6434; \citealt{Sun2025SAPPHRES}). Bright wisps are rare, occurring in $\sim$ 1\% of current observations but are significantly more luminous than normal wisps. Their presence in the LW channel suggests a possibly different origin, likely associated with a very specific gap in the NIRCam light path. Unlike normal wisps, their occurrence and morphology depend sensitively on telescope pointing and position angles, although the triggering conditions remain unclear. In the SAPPHIRES program, they appear in one observation but not another that observes the same field at similar position angles but several months later. 

Bright wisps have significantly different morphologies, showing more elongated and complex sharp stripes. The morphology variations between exposures are dramatic. We have not collected sufficient data to perform multi-component decomposition. \cite{Duncan2025JELS} effectively remove the LW scattered light using a median-stacked template but discard wisp-contaminated regions in SW images. The method of \cite{robotham_2023_wisp} may be suited for subtracting bright wisps in SW images, although careful treatment is required for associated diffuse emission.

\subsection{Claws: a distinct scattered light phenomenon}
Claws are another type of scattered light artifact  with a claw-mark-like appearance in JWST NIRCam images. They are caused by bright stars and observed only in the SW channel. Different from wisps, the appearances of claws are sensitive to the relative position between the star and the field of view. They are not always present in an image and can appear at different locations on the detector. Claws cannot be subtracted using templates due to their large variation in morphologies and locations. The best practice is therefore to mask them manually. They are easily identified because they are sharp and elongated, an appearance distinct from sources and wisps.

\bigskip

\section{Summary}
\label{sec:summary}
Wisps in JWST/NIRCam imaging introduce spatially correlated, low-surface-brightness systematics that bias photometry and morphology, even after image stacking. They are a major source of contamination for faint-source measurements critical to studies such as high-redshift galaxy identification. We develop a NMF framework that models wisp contamination using a small set of non-negative templates, capturing wisp morphologies and their exposure-to-exposure variations. Applied to the deep JADES observations, this method produces clear visual improvements in wisp subtraction. Consequently, photometric bias and scatter for sources in wisp-contaminated regions are reduced to levels nearly consistent with those in unaffected regions, and large-scale mosaic artifacts are substantially suppressed. We publicly release the wisp templates, subtraction code, and template-construction pipeline, which are readily applicable to future programs.

\begin{acknowledgments}
This work is based on observations made with the NASA/ESA/CSA James Webb Space Telescope. The data were obtained from the Mikulski Archive for Space Telescopes at the Space Telescope Science Institute, which is operated by the Association of Universities for Research in Astronomy, Inc., under NASA contract NAS 5-03127 for JWST. These observations are associated with JWST programs 1176, 1180, 1181, 1210, 1264, 1283, 1286, 1287, 1895, 1963, 2079, 2198, 2514, 2516, 2674, 3215, 3577, 3990, 4540, 4762, 5398, 5997, 6434, 6511, and 6541. The authors acknowledge the teams of programs 1895, 1963, 2079, 2514, 3215, 3577, 3990, 6434, and 6541 for developing their observing program with a zero-exclusive-access period. 

This research was funded through JWST program 3905. Support for program 3905 was provided by NASA through a grant from the Space Telescope Science Institute, which is operated by the Association of Universities for Research in Astronomy, Inc., under NASA contract NAS 5-03127. BDJ, DJE, PC, KH, BER, and CNAW acknowledge support from the NIRCam Science Team contract to the University of Arizona, NAS5-02015. DJE is also supported as a Simons Investigator and by NASA through a grant from the Space Telescope Science Institute, which is operated by the Association of Universities for Research in Astronomy, Inc., under NASA contract NAS5-03127. RH acknowledges funding provided by the Johns Hopkins University, Institute for Data Intensive Engineering and Science (IDIES). BER also acknowledges support from JWST Program 3215. ST acknowledges support by the Royal Society Research Grant G125142. The research of CCW is supported by NOIRLab, which is managed by the Association of Universities for Research in Astronomy (AURA) under a cooperative agreement with the National Science Foundation. This research made use of the lux supercomputer at UC Santa Cruz which is funded by NSF MRI grant AST 1828315.

\end{acknowledgments}

\appendix
\restartappendixnumbering 
\twocolumngrid
\section{Sky Flat Modeling}
\label{sec:skyflat}
NIRCam shows pixel-to-pixel sensitivity variations that are corrected by flat-field calibration with flat images. If the true flat differs from the estimated flat, residual flat-field errors can imprint spurious spatial structure on the images via the sky background, an effect that is most significant for background-dominated observations. We evaluate the flat-field errors of the STScI flat image in the deep JADES observations using the NMF method.

First, we illustrate how inaccurate flat field produces spurious large-scale  low-surface-brightness artifacts.  Let the raw ramp data be 
\begin{equation}
    R(x)=F(x)[S(x) + B],
\end{equation}
where $F(x)$ is the true flat-field response, $S(x)$ is the astrophysical signal, $B$ is the sky background as a constant, and $x$ is the pixel coordinates. After flat-field correction using an estimated flat $F'(x)$, which differs from the true flat by $\epsilon(x)\equiv F(x)/F'(x)-1$, and after subtracting a constant background $B'$, the calibrated data $D(x)$ are obtained as
\begin{equation}
\begin{aligned}
    D(x) &= R(x)/F'(x)- B' \\&=(1+\epsilon(x))[S(x) + B] - B' \\
    &= (1+\epsilon(x))S(x)+ \epsilon(x) B + (B-B') \\
    &\simeq S(x)+ \epsilon(x) B,
\end{aligned}
\end{equation}
where we assume $\epsilon(x)\ll1$. 
The term $\epsilon(x) B$ is not negligible because the sky background $B$ can be large. The sky background is around $0.09\,\mathrm{MJy\,sr^{-1}}$ in the JADES observations. Therefore, a detector patch with a $1\%$ flat error may produce a low-surface-brightness feature with a brightness of $\sim10^{-3}\,\mathrm{MJy\,sr^{-1}}$ (28\,mag\,arcsec$^{-2}$).

\begin{figure}
    \centering
    \includegraphics[width=\linewidth]{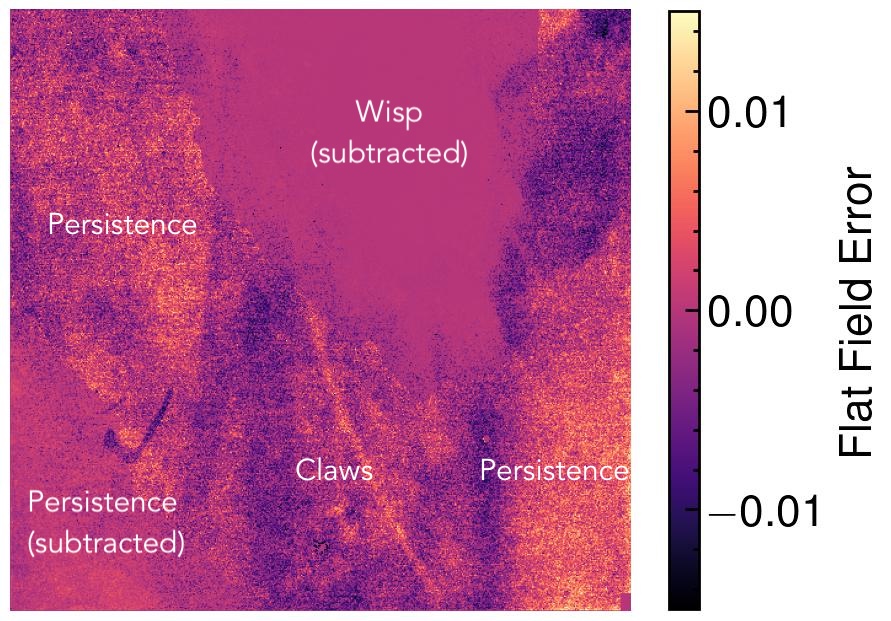}
    \caption{Inferred flat-field errors for the NIRCam B4 detector derived from deep JADES observations using the NMF framework. The map shows the flat-field error, $\epsilon= F/F' - 1$, where $F$ is inferred on-sky flat, and $F'$ is the STScI flat. Regions affected by known persistence and claw contamination are annotated. Outside these contaminated areas, the fluctuations indicate a flat-field accuracy at the $\lesssim 1\%$ level.}
    \label{fig:flat}
\end{figure}

We estimate the flat field errors with the NMF algorithm.  To simultaneously model the flat field $F_n$, the background $B_m$, and the wisps $W_{kn}$ with amplitudes $A_{mk}$, we formulate the rate image ${R}_{mn}$ as the sum of a wisp term and a sky background term:
\begin{align}
{R}_{mn} = \left(\sum_k A_{mk}W_{kn}\right) F_n + B_mF_n.
\end{align}
where $m$ is the number of images, $n$ is the number of pixels in each image, and $k$ is the number of wisp templates. To reconcile with previous equations, we define the normalized data and flat as
\begin{align}
D_{mn} \equiv R_{mn}/F_n, \qquad
M_{n} \equiv F_n/F'_n,
\end{align}
where the divisions are performed element-wise.  It is convenient to absorb the residual flat-field term into the wisp template by defining ${W}'_{kn} \equiv {W}_{kn} M_n$. We obtain
\begin{align}
{D}_{mn} = \sum_k {A_{mk}}{W'_{kn}} + {B_m}M_n.
\end{align}

The flat and the wisps are degenerate since they are permutation-symmetric in the equation. The degeneracy is mitigated by setting wisps to zero outside the wisp region, while fitting the flat field across the whole detector. We also impose priors that keep the flat field close to unity, as the STScI flat should be close the true flat. With a parameter $\alpha$ controls the prior strength, the loss function is
\begin{equation}
\begin{aligned}
    \mathcal{L} = \sum_{i,j}\Omega_{ij} \left( D_{ij} - \sum_k A_{ik} W'_{kj} - B_i M_j \right)^2\\
    + \sum_j \alpha_j^2 \left(M_j - 1\right)^2 .
\end{aligned}
\end{equation}

The derivatives with respect to $B_i$, $M_j$, $W'_{kj}$, and $A_{ik}$ are
\begin{equation}
\frac{\partial L}{\partial W'_{kj}} =
-2 \sum_i \Omega_{ij} A_{ik} \left( D_{ij} - \sum_k A_{ik} W'_{kj} - B_i M_j \right), 
\end{equation}
\begin{equation}
\begin{aligned}
\frac{\partial L}{\partial M_{j}} =
-2 \sum_i \Omega_{ij} B_i \left( D_{ij} - \sum_k A_{ik} W'_{kj} - B_i M_j \right) \\
+ 2 {\alpha_j^2} \left(M_j - 1\right), 
\end{aligned}
\end{equation}
\begin{equation}
\frac{\partial L}{\partial A_{ik}} =
-2 \sum_j \Omega_{ij} W'_{kj} \left( D_{ij} - \sum_k A_{ik} W'_{kj} - B_i M_j \right),
\end{equation}
\begin{equation}
\frac{\partial L}{\partial B_{i}} =
-2 \sum_j \Omega_{ij} M_j \left( D_{ij} - \sum_k A_{ik} W'_{kj} - B_i M_j \right).
\end{equation}

The multiplicative updates are
\begin{equation}
\begin{aligned}
W'_{kj} &\leftarrow W'_{kj} \times
\frac{\sum_i \Omega_{ij} A_{ik} D_{ij}}
{\sum_i \Omega_{ij} A_{ik} (\mathbf{A}\mathbf{W'})_{ij}
+ \sum_i \Omega_{ij} A_{ik} B_i M_j},\\
M_j &\leftarrow M_j \times
\frac{\sum_i \Omega_{ij} B_i D_{ij} + \alpha_j^2}
{\sum_i \Omega_{ij} B_i (\mathbf{A}\mathbf{W'})_{ij}
+ \sum_i \Omega_{ij} B_i^2 M_j
+ \alpha_j^2 M_j}.
\end{aligned}
\end{equation}

and
\begin{align}
A_{ik} &\leftarrow A_{ik} \times
\frac{\sum_j \Omega_{ij} W'_{kj} D_{ij}}
{\sum_j \Omega_{ij} W'_{kj} (\mathbf{A}\mathbf{W'})_{ij}
+ \sum_j \Omega_{ij} W'_{kj} B_i M_j}, \\
B_i &\leftarrow B_i \times
\frac{\sum_j \Omega_{ij} M_j D_{ij}}
{\sum_j \Omega_{ij} M_j (\mathbf{A}\mathbf{W'})_{ij}
+ \sum_j \Omega_{ij} M_j^2 B_i}.
\end{align}

The prior strength $\alpha_j$ is applied only to pixels within the wisp regions, where it is set to $100\,m$, with $m=379$ exposures. Outside the wisp regions we set $\alpha_j=0$. We also simultaneously fit the morphology and amplitudes of a persistence feature in the lower-left corner of the B4 detector using the NMF algorithm.

Figure~\ref{fig:flat} shows the inferred flat-field error, defined as $\epsilon = M - 1=F/F'-1$. Because persistence effects are degenerate with flat-field errors, the image contains imprints from persistence. Outside regions affected by known persistence, the inferred flat-field error is of order $\sim$ 1\%. Errors at this level would produce background artifacts fainter than 28\,mag\,arcsec$^{-2}$ and induce spurious source variability at the $\sim$ 1\% level when the same source is observed at different locations on the detector.

\bibliographystyle{aasjournal}
\bibliography{reference}
\end{document}